\documentclass[reprint, amsmath,amssymb,aps,pre,notitlepage,superscriptaddress,twocolumn
]{revtex4-1}
\usepackage{blindtext}
\usepackage{comment}
\usepackage{tikz}
\usepackage{subfigure}
\usepackage{graphicx}
\usepackage{dcolumn}
\usepackage{mathtools,amsmath}
\usepackage{amssymb}
\usepackage{hyperref}
\usepackage{float}
\usepackage{graphicx}
\usepackage{subfigure,bm}
\usepackage[export]{adjustbox}
\usepackage{type1cm}
\usepackage{eso-pic}
\usepackage{color}
\usepackage{dcolumn}
\newcolumntype{x}[1]{>{\centering\arraybackslash\hspace{0pt}}p{#1}}
\newcommand{\nnt}{{\bm n} \otimes {\bm n}}

\begin{document}

\preprint{APS/123-QED}
\title{Rheology of dense granular suspensions under extensional flow}
\author{Oliver Cheal}
\affiliation{Cavendish Laboratory, University of Cambridge, Cambridge CB3 0HE, United Kingdom}
\author{Christopher Ness}
\affiliation{Department of Chemical Engineering and Biotechnology, University of Cambridge, Cambridge CB3 0AS, United Kingdom}
\pacs{Valid PACS appear here}
\date{\today}

\begin{abstract}
We study granular suspensions under a variety of extensional deformations and simple shear using numerical simulations.
The viscosity and Trouton's ratio (the ratio of extensional to shear viscosity) are computed as functions of solids volume fraction $\phi$ close to the limit of zero inertia.
Suspensions of frictionless particles follow a Newtonian Trouton's ratio for $\phi$ all the way up to $\phi_0$,
a universal jamming point that is independent of deformation type.
In contrast, frictional particles lead to a deformation-type-dependent jamming fraction $\phi_m$,
which is largest for shear flows.
Trouton's ratio consequently starts off Newtonian but diverges as $\phi\to\phi_m$.
We explain this discrepancy in suspensions of frictional particles by considering the particle arrangements at jamming.
While frictionless particle suspensions have a nearly isotropic microstructure at jamming,
friction permits more anisotropic contact chains that allow jamming at lower $\phi$ but introduce protocol dependence.
Finally, we provide evidence that \emph{viscous number} rheology can be extended from shear to extensional deformations, with a particularly successful collapse for frictionless particles.
Extensional deformations are an important class of rheometric flow in suspensions, relevant to paste processing, granulation and high performance materials.
\end{abstract}

\maketitle

\section{Introduction}

Industrial and geophysical processes that involve dense suspensions in motion invariably exhibit combinations of shear and extensional flow~\cite{benbow1993paste,macosko1994rheology}.
To achieve a useful description of their rheological properties,
one must therefore start with a sound knowledge of the material response to both types of deformation.
Despite this clear requirement,
most of the recent influential developments in the understanding of suspension rheology (both experimental~\cite{boyer2011unifying,Guy2015} and numerical~\cite{trulsson2012transition,andreotti2012shear,lerner2012unified}),
and indeed dry granular material rheology (see for example Refs~\cite{chialvo2012bridging,jop2006constitutive}),
have focussed exclusively on shear flows. 
This shortcoming is understandable in part due to the relative difficulty of achieving purely extensional flows experimentally.
Extensional deformations are, however, more severe than shearing in the sense that material elements move apart exponentially, rather than linearly, with time (or strain)~\cite{brader2008first}, so in practical applications they may well prove to dominate the overall rheological phenomenology. The relative importance of extensional to shear rheological properties is traditionally quantified using Trouton's ratio, the ratio of extensional to shear viscosity.

Extensional rheology is better understood in polymers,
with many successful experimental approaches having been developed over the past four decades.
Classical techniques include melt stretching~\cite{meissner1969rheometer}, filament stretching~\cite{sridhar1991measurement,mckinley2002filament}, flows through a contraction~\cite{binding1988use,everage1978extensional,mackley1995multipass,oliveira2007viscous,oliveira2008simulations,rothstein1999extensional}
and lubricated squeezing~\cite{chatraei1981lubricated,binding1988approximate}.
Led by experimental insights from such techniques,
constitutive models for polymer rheology have long benefitted from understanding both shear and extensional flows~\cite{Rubinstein2003,tripathi2006rheology}.
For the continued progression of the field of suspension rheology,
it is essential that the understanding of arbitrary deformations can be brought up to speed with that of polymers.

In recent years, there have been a number of experimental studies
of the extensional rheology of dense suspensions using similar techniques to those above.
There has been notable emphasis on shear thickening systems, a particular class of suspension that sits close to the colloidal-granular interface~\cite{Guy2015}.
A popular route has been to use a filament stretching device to probe the high deformation rate uniaxial extension regime.
In such experiments, Rothstein and coworkers observed strain hardening in nano- and micrometre particle suspensions,
with light scattering results suggesting particle self-organisation as the origin~\cite{chellamuthu2009extensional,white2010extensional}.
The approach is robust enough to detect changes in particle concentration and solvent properties~\cite{khandavalli2014extensional} and to examine properties relevant to printing and other applications~\cite{khandavalli2017ink,tiwari2009elongational}.
Devices of this type have the added complexity of a liquid-air interface,
the shape of which distorts under rapid extensional flows,
leading naturally to a connection between strain hardening and granule formation~\cite{smith2010dilatancy}.
Another series of experiments placed a tensile load on a cornstarch suspension~\cite{majmudar2017dynamic},
leading to the unexpected result that shear jamming and shear thickening,
both purportedly manifestations of stress-induced particle friction~\cite{wyart2014discontinuous},
can be independently inhibited using chemical modifiers~\cite{james2017interparticle}.

A simulation model predicting shear thickening under extensional flow has emerged recently~\cite{seto2017microstructure}, and found a Trouton's ratio of 4 for planar extension (both above and below shear thickening), surprisingly consistent with the prediction for a Newtonian fluid. The analysis focussed on the effect of polydispersity on shear thickening at a small number of volume fractions. It is has not yet been explored whether there is solids volume fraction dependence or deformation-type dependence beyond planar extension.
Experimental measurements of Trouton's ratio have been reported for suspensions  and found to be $\mathcal{O}(10)$ in the Brownian regime~\cite{xu2005shear} and slightly above the expected Newtonian values in the granular regime~\cite{dai2017elongational}. Furthermore, particle roughness was found to enhance the extensional viscosity, demonstrating the importance of including particle-particle friction in numerical models and constitutive descriptions of extensional flow.

The focus on shear thickening is understandable given its ubiquity in applications,
but it has led to an overlooking of the underlying rheological behaviour of granular (by which we mean athermal) suspensions under extensional flow.
This underlying behaviour is typically described under shear flow by the much-used viscous number model, the so-called $\mu(I_v)$-rheology~\cite{boyer2011unifying}.
It is a robust framework for rheological modelling of granular particles of arbitrary particle-particle friction,
and takes as its basis the assumption that for sufficiently hard spheres the only relevant stress scale is the hydrodynamic one.
This leads naturally to apparent Newtonian rheology in which all stresses scale linearly with shear rates, but remain highly sensitive to the solid volume fraction $\phi$.
The presence of a particle pressure under shear flow~\cite{boyer2011unifying},
complicated by an ambiguity in measured values of reported normal stress differences~\cite{royer2016rheological,pan2017normal}, however, leads crucially to the denomination `quasi'-Newtonian for any dense granular suspension described by $\mu(I_v)$-rheology.
Consequently, it is not clear whether this framework can be generalised to extensional flows,
and in particular whether ratios of normal to shear stresses (expressed as Trouton's ratio) should be truly Newtonian in such circumstances,
in spite of the direct proportionality between stresses $\sigma$ and shear rates $\dot{\gamma}$.
In this respect, extensional flows are an important class of rheometric flow for studying suspensions close to jamming.

In this article, we use numerical simulations to study dense granular suspensions under extensional flow.
We implement a minimal discrete element-type numerical model
that keeps track of the trajectories and forces on overdamped,
neutrally buoyant suspended spheres,
which are updated in a deterministic way according to Newtonian dynamics.
The force terms comprise hydrodynamic lubrication and harmonic contact potentials with friction. 
The model operates in the athermal, non-inertial regime.
We consider planar, uniaxial and biaxial deformations and compute the Trouton's ratios as a function of the solid volume fraction,
using shear flow as a reference.
The distinction identified recently between sliding and rolling contacts for suspended particles~
\cite{fernandez2013microscopic,comtet2017pairwise} and, by extension,
the role of frictional forces in suspensions of large particles,
ought still to be valid for extensional flows,
so it is essential to consider explicitly the role of static friction between particles.
For this reason, our model allows hydrodynamic lubrication forces to break down on some surface roughness lengthscale,
and we thereafter consider direct particle-particle contacts with static friction coefficient $\mu$.

We first describe our numerical simulation methodology (Section~\ref{section_model}),
before describing the deformation types studied and the method of imposing them (Section~\ref{section_deformations}).
We then describe the response of the material during the straining period (Section~\ref{section_evolution}),
and go on to demonstrate the divergence of the shear and extensional viscosities with volume fraction (Section~\ref{section_divergence}).
We find a discrepancy in the critical volume fractions for suspensions of frictional particles that can be explained by considering the microstructural configurations at jamming (Section~\ref{section_microstructure}). Finally, we discuss to what extent the results for extensional flow can be mapped on to viscous number rheology (Section~\ref{section_viscous}).

\section{Numerical model}
\label{section_model}

Our model considers athermal, noninertial, neutrally bouyant particles that represent a suspension corresponding to that used in the seminal experiment of~\citet{boyer2011unifying}.
We consider a periodic domain containing bidisperse spheres with solids volume fraction $\phi$. The particles have density $\rho$ and radii $a$ and $1.4a$, mixed in equal numbers. The simulation box is initialised with 12,000 (shear) or 15,000 (extension) particles (we explore the importance of system size in the Appendix) placed randomly before being relaxed to achieve minimally overlapping states. In what follows, we report ensemble averages over five realisations obtained by changing the initial configurations using a random seed.
The simulation box is deformed according to a velocity gradient tensor ${\bm U}^\infty$. Suspended particles are thus subjected to a rate of strain tensor with symmetric and antisymmetric parts ${\textbf E}^\infty$ and ${\bm \Omega}^\infty$ respectively, where the background fluid flow at ${\bm x}$ follows ${\bm U}^\infty({\bm x}) = {\textbf E}^\infty{\bm x} + {\bm \Omega}^\infty \times {\bm x}$.

\paragraph{Hydrodynamic forces}
Hydrodynamic interactions between particles are based upon the resistance matrix formalism described by Refs~\cite{jeffrey1984calculation,jeffrey1992calculation,kim1991microhydrodynamics}. Following~\citet{ball1997simulation} we consider short-ranged, frame-invariant, pairwise interactions. For neighbouring particles 1 and 2, translating with velocities ${\bm U}_1$, ${\bm U}_2$ and rotating at ${\bm \Omega}_1$, ${\bm \Omega}_2$, and with centre-centre vector ${\bm r}$ (and ${\bm n}  = {\bm r}/|{\bm r}|$) pointing from particle 2 to particle 1, it can be shown~\cite{Ranga2017} that the force ${\bm F}^h$ and torque ${\bm \Gamma}^h$ on particle 1 are given by:
\begin{subequations}
\begin{equation}
\begin{split}
{\bm F}^h/\eta_f =& (X^A_{11} \nnt + Y_{11}^A({\bm I}-\nnt))({\bm U}_2 - {\bm U}_1)\\
&+ Y^B_{11}({\bm \Omega}_1 \times {\bm n})+Y^B_{21}({\bm \Omega}_2 \times {\bm n}) \text{,}
\end{split}
\end{equation}
\begin{equation}
\begin{split}
{\bm \Gamma}^h/\eta_f =& Y^B_{11}({\bm U}_2 - {\bm U}_1)\times {\bm n}\\
&- ({\bm I}-\nnt)(Y_{11}^C{\bm \Omega}_1 + Y_{12}^C{\bm \Omega}_2) \text{,}
\end{split}
\end{equation}
\end{subequations}
where $\eta_f$ is the Newtonian viscosity of the suspending liquid.
For particle radii $a_1$ and $a_2$, the surface-surface separation is given by $h = |{\bm r}| - (a_1 + a_2)$, which we nondimensionalise as $\xi = 2h/(a_1+a_2)$. The scalar resistances $X^A_{11}$, $Y^A_{11}$, $Y^B_{11}$, $Y^B_{21}$, $Y^C_{11}$ and $Y^C_{12}$ comprise short range contributions that diverge as $1/\xi$ and $\ln(1/\xi)$ and are given in Appendix A. We neglect interactions that have $h>0.05a$ (with $a$ the smaller particle radius). The per-force hydrodynamic stresslet is ${S}_{ij}^h =  -\frac{1}{2}\left({F}_i^h {r}_j + F_j^hr_i\right)$.
A drag force and torque act on particle 1 at position ${\bm x}_1$, given by 
\begin{subequations}
\begin{equation}
{\bm F}^{d} = -6\pi \eta_fa_1({\bm U}_1 - {\bm U}^\infty({\bm x}_1)) \text{,}
\end{equation}
\begin{equation}
{\bm \Gamma}^{d} = -8\pi \eta_f a_1^3({\bm \Omega}_1 - {\bm \Omega}^\infty({\bm x}_1)) \text{,}
\end{equation}
\end{subequations}
leading to per-particle contributions to the stresslet given by ${\bm S}^{d} = -\frac{20\pi}{3}\eta_fa_1^3{\textbf E}^\infty$.

\paragraph{Contact forces}
Following experimental evidence that lubrication layers break down in suspensions under large stress~\cite{fernandez2013microscopic}, and, equivalently, for large particles~\cite{Guy2015}, we use a minimum $h$ defined as $h_\mathrm{min}=0.001a$, below which hydrodynamic forces are regularised and particles may come into mechanical contact. For a particle pair with contact overlap $\delta = ((a_1+a_2)-|{\bm r}|)\Theta((a_1+a_2)-|{\bm r}|)$ and centre-centre unit vector ${\bm n}$, we compute the contact force and torque on particle 1 according to \cite{cundall1979discrete}:
\begin{subequations}
\begin{equation}
{\bm F}^c = k_n\delta{\bm n} - k_t{\bm u}
\end{equation}    
\begin{equation}
{\bm \Gamma}^c = a_1 k_t({\bm n} \times {\bm u})
\end{equation}
\end{subequations}
where ${\bm u}$ represents the incremental tangential displacement, reset at the initiation of each contact. $k_n$ and $k_t$ are stiffnesses, with $k_t = (2/7)k_n$. The tangential force component is restricted by a Coulomb friction coefficient $\mu$ such that $|k_t{\bm u}| \leq \mu k_n\delta$. For larger values of $|k_t{\bm u}|$, contacts enter a sliding regime. We take the stresslet as ${S}_{ij}^c =  -{F}_i^c {r}_j$.

Trajectories are computed from the above forces. Contact and hydrodynamic forces and torques are summed on each particle and the trajectory is updated according to a Velocity-Verlet algorithm. The dynamics are controlled by three dimensionless quantities: the volume fraction $\phi$, the Stokes number $\text{St}$ and the stiffness-scaled shear rate $\hat{\dot{\gamma}}$.
We ensure that the Stokes number $\text{St}=\rho\dot{\gamma}a^2/\eta_f$ remains $\ll1$ throughout to approximate over-damped conditions. We found $\mathcal{O}(10^{-3})$ to be sufficiently small in practice and achieved this by setting particle radius $a=0.5$ [$\emph{length}$], density $\rho=1$ [$\emph{mass}/\emph{length}^3$], suspending fluid viscosity $\eta_f=0.1$ [$\emph{mass}/(\emph{time}\times\emph{length})$] and shear rate $\dot{\gamma}=0.001$ [\emph{1/time}]. In this limit we expect rate-independent rheology in which all stresses scale linearly with deformation rates. The extent to which the particles may be considered hard spheres is set by the shear rate rescaled with particle stiffness, as given by $\hat{\dot{\gamma}} = 2\dot{\gamma}a/\sqrt{k_n/(2 \rho a)}$~\cite{chialvo2012bridging}. We set $\hat{\dot{\gamma}}<10^{-5}$ throughout by setting $k_n=50000$ [$\emph{mass}/\emph{time}^2$].
The model is implemented in $\texttt{LAMMPS}$~\cite{plimpton1995fast}. The overall stress tensor is computed by summing over all of the stresslets
$
{\bm \sigma} =  -2 \eta_f {\textbf E}^\infty +   \frac{1}{V}\left(\sum_i {\bm S}^d + \sum_{p_h} {\bm S}^h + \sum_{p_c} {\bm S}^c \right)
$
where the sums are over individual particles $i$, hydrodynamically interacting pairs $p_h$ and contacting pairs $p_c$.

\section{Description of applied deformations}
\label{section_deformations}

\begin{center}
\begin{table}
\begin{tabular}{x{13mm}  x{24mm}  x{16mm}  x{27mm} }
 & Rate of deformation ${\textbf E}^\infty$ & Magnitude $|{\textbf{E}}^\infty|$ & Viscosity \\ \\ \hline \\
Simple shear
  &
  $\begin{pmatrix}
    0 &\frac{1}{2}\dot{\gamma} & 0  \\
    \frac{1}{2}\dot{\gamma} & 0                    & 0 \\
    0 & 0                    & 0 
  \end{pmatrix}$
  & 
  $\dot{\gamma}$
  &$\sigma_{12}/\dot{\gamma} = \eta^\dagger$ \\ \\
  Planar extension
  &
  $\begin{pmatrix}
    -\dot{\gamma} & 0 & 0  \\
    0 &  \dot{\gamma}    & 0 \\
    0 & 0 &  0
  \end{pmatrix}$
  & 
  $2\dot{\gamma}$
  &$(\sigma_{22} - \sigma_{11})/ \dot{\gamma}  = 4\eta^\dagger$\\ \\
  Uniaxial extension
  &
  $  \begin{pmatrix}
    -\frac{1}{2}\dot{\gamma} & 0 & 0  \\
    0 &   -\frac{1}{2}\dot{\gamma}   & 0 \\
    0 & 0 &  \dot{\gamma}
  \end{pmatrix}$
  & 
  $\sqrt{3}\dot{\gamma}$
  &$(\sigma_{33} - \sigma_{11})/\dot{\gamma}  =  3\eta^\dagger$    \\\\
  Biaxial extension
  &
  $ \begin{pmatrix}
    \dot{\gamma} & 0 & 0  \\
    0 &  \dot{\gamma}    & 0 \\
    0 & 0 &  -2\dot{\gamma}
  \end{pmatrix}$
  &
  $2\sqrt{3}\dot{\gamma}$
  & $(\sigma_{11} - \sigma_{33})/ \dot{\gamma}  = 6\eta^\dagger$ \\ \\ \hline
  \end{tabular}
      \caption{Rate of deformation tensors ${\textbf E}^\infty$, their magnitudes, and the viscosity definitions for each type of flow explored in this work.
      }
      \label{table1}
  \end{table}
\end{center}

\begin{figure*}
  \centering
  \includegraphics[trim = 0mm 0mm 87mm 0mm, clip,width=0.95\textwidth]{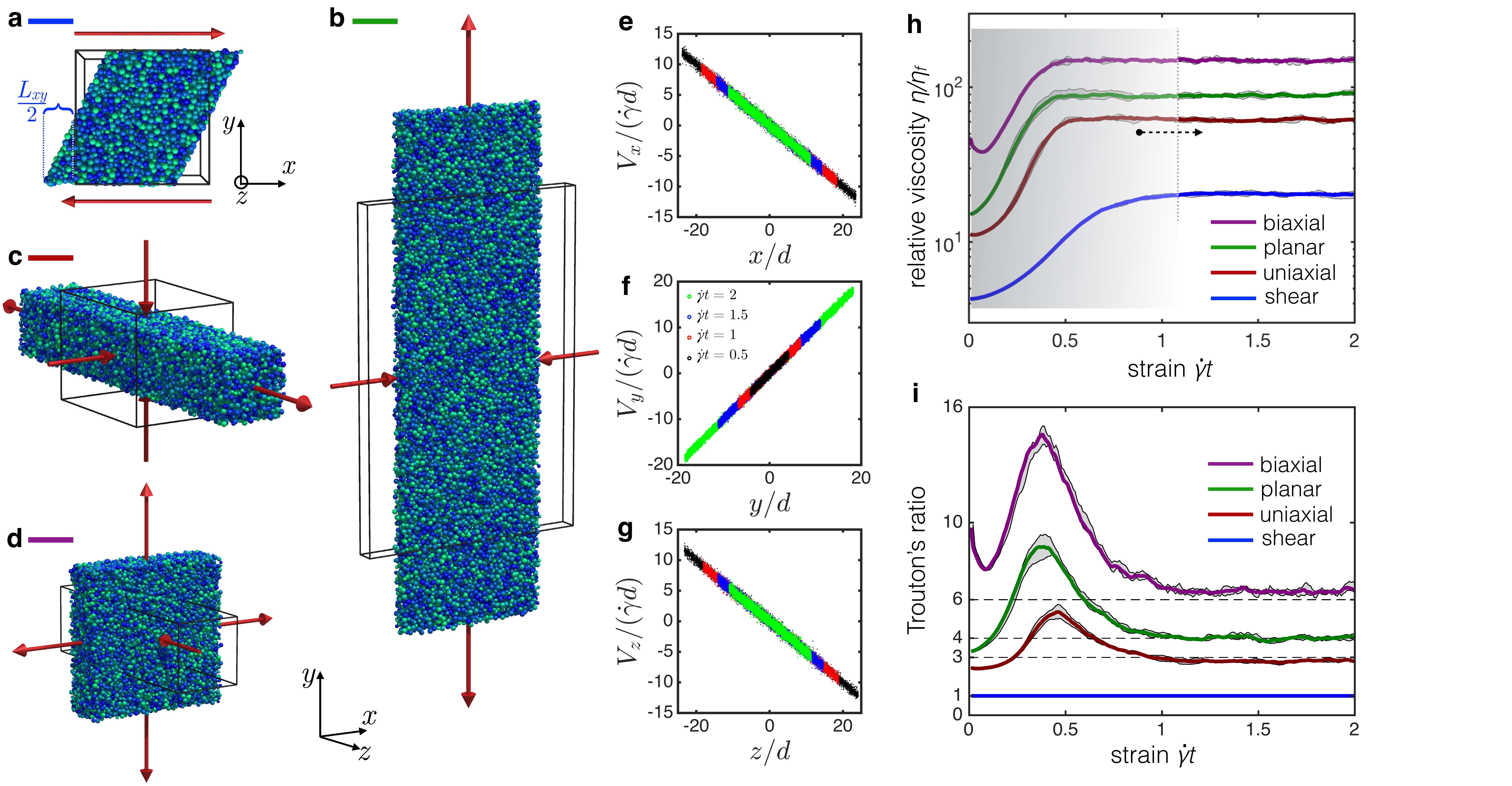}
     \caption{
Schematics of the deformations applied in this work. Shown are
(a) simple shear;
(b) planar extension;
(c) uniaxial extension;
(d) biaxial extension.
In each case the wireframe box illustrates the box dimensions at an earlier time and the red arrows indicate the directions of the applied deformation. The upper coordinate diagram refers to (a) while the lower one refers to (c)-(d).
The box deformations lead to uniform velocity gradients. Shown in (e)-(g) are examples of the velocity gradients obtained during uniaxial extension at increasing strain increments.
(h) Plot of viscosity as a function of strain for each flow type (at $\phi=0.45$ and friction $\mu=1$) showing start-up period (shaded) and the steady flow period (unshaded). Black arrow indicates region from which viscosities are used for averaging.
(i) Plot of Trouton's ratio as a function of strain for each flow type. Dashed lines represent the corresponding Newtonian values.
In each case the grey shaded area represents the maximum and minimum values obtained during five independent simulation runs.
Coloured bars next to figure labels (a)-(d) correspond to the colours in (h) and (i).
     }
 \label{figure1}
\end{figure*}

We consider shear, planar, uniaxial and biaxial flows
as illustrated schematically in Figures~\ref{figure1}(a)-(d) respectively.
Consider the general velocity gradient tensor ${\bm U}^\infty$ in three dimensions
which has components $\partial v_i/\partial x_j$.
From this, we obtain the components of the symmetric rate of deformation tensor as
$
E^\infty_{ij} = \frac{1}{2}\left(\frac{\partial v_i}{\partial x_j} + \frac{\partial v_j}{\partial x_i}\right) \text{.}
$
Shear flows have a corresponding rotational part
$
\Omega^\infty_{ij} = \frac{1}{2}\left(\frac{\partial v_i}{\partial x_j} - \frac{\partial v_j}{\partial x_i}\right)
$,
while extensional flows are irrotational.
For the case of a Newtonian fluid, the stress tensor is then given simply by
$\sigma_{ij} = -p\delta_{ij} + 2\eta^\dagger E^\infty_{ij}$
where $\eta^\dagger$ is the Newtonian viscosity.
In Table~\ref{table1} we present the rate of deformation tensors ${\textbf E}^\infty$
corresponding to each of the flow types explored in this work,
as well as the magnitudes $|{\textbf{E}}^\infty| = \sqrt{2 {{\textbf E}^\infty}:{{\textbf E}^\infty}}$ and the associated viscosity definitions, which follow Ref~\cite{macosko1994rheology}.
It is noted that the framework of~\citet{jones1987extensional} dictates that for the uniaxial Trouton's ratio we should compare the shear viscosity at $\dot{\gamma}$ with the uniaxial extensional viscosity at $\sqrt{3}\dot{\gamma}$. Comparing the reported deformation rate magnitudes in Table~\ref{table1}, we see that this requirement is satisfied.
We take the Trouton's ratios (Tr) as the ratios of each of the extensional viscosities to the shear viscosity.
These lead to values of the Newtonian Trouton's ratio of 4, 3 and 6 for planar, uniaxial and biaxial flows, respectively.
We will use these values as a basis for comparison for the extensional flows modelled in this work.

Volume-conserving deformations are applied to the simulated suspension
by incrementally changing the dimensions of the periodic box according to the relevant rate of deformation tensor.
To simulate simple shear, we use a triclinic periodic box with a tilt length $L_{xy}$ (see Figure~\ref{figure1}(a)) that is incrementally
increased linearly in time as $L_{xy}(t) = L_{xy}(t_0) + L_y \dot{\gamma}t$,
giving a deformation that is entirely equivalent to that obtained using a Lees-Edwards boundary condition.
For extensional flows, the leading box dimension is increased with time according to
$L(t)=L(t_0)e^{\dot{\gamma}t}$ to give a constant true strain rate,
that we quantify as $\dot{\gamma}$.
The other box dimensions are varied accordingly to conserve the volume.
We verified that neglecting particle-particle interactions and imposing simply the deformation protocol described here and the Stokes drag forces described above leads to particle trajectories that follow precisely the affine deformation of the simulation box.
The velocity of any particle that crosses a periodic boundary is remapped according to the velocity gradient across the box perpendicular to that boundary.
The velocity gradient at any point in the simulation box at any time represents the overall applied box deformation and thus the particles are subjected to uniform velocity gradients as illustrated in Figure~\ref{figure1}(e)-(g).

Whereas the shear deformation can be continually remapped to permit arbitrarily large deformations, the extensional deformations are constrained in magnitude 
since our simulation approach involves `shrinking' one of the box dimensions with time.
Taking the uniaxial deformation as an illustrative example,
we initiate the simulation box with 15,000 particles of radii $a$ and $1.4a$ and with a cuboidal
box of dimensions $171.4a \times171.4a \times 10a$ (giving $\phi = 0.4$ in this case).
During the period for which we observe strain-independence of the viscosity (see below),
the box dimensions remain $\mathcal{O}(10)a$ in $x$, $y$ and $z$.
There is uniform straining throughout the sample during this period, with steady state locally acting velocity gradients that match the overall box deformation (see Figure~\ref{figure1} and Supplementary Video).
We verified that there is no system size dependence by simulating a smaller sample
and achieving a comparable (though shorter) steady-state period.
Our numerical model breaks down at large extensional strains as the contracting dimensions of the
simulation box reach $\mathcal{O}(1)$ particle radii and particles `see' themselves through periodic images (see Appendix).
Notwithstanding the difficulty in achieving large deformations, the approach we describe here to achieve steady velocity gradients during various extensional deformations has been discussed and applied previously in several works across glassy and polymeric systems (see, for example, Refs~\cite{hounkonnou1992liquid,todd1997elongational,todd1998nonequilibrium,ruiz2017simulations,soules1983rheological,lavine2003molecular,lyulin2004molecular}).
To reach larger strains, it is necessary to implement remappings such as those described by Kraynik and Reinelt~\cite{kraynik1992extensional,hunt2016periodic,seto2017microstructure} for planar deformations. For materials involving long time/length scales (polymer melts for instance), these boundaries are essential. For dense suspensions, one the other hand, that can reach a steady state within strains of 1 or 2~\cite{Gadala-maria1980}, they may be useful for some studies but are not crucial to study steady phenomena. The planar deformation used in this work is equivalent to that acting between remappings of the Kraynik-Reinelt scheme.

\section{Evolution of suspension viscosity with strain}
\label{section_evolution}

Starting from a quiescent state with minimal particle-particle contacts, we begin the constant-rate deformation.
The viscosity for each is computed from the simulation data according to the definitions in Table~\ref{table1}.
For example, the suspension viscosity under uniaxial extension is given by $\eta = (\sigma_{zz}-\sigma_{xx})/\dot{\gamma}$.
This result is further rescaled by the suspending fluid viscosity $\eta_f$ and we thus present reduced viscosities as $\eta/\eta_f$. 
Viscosity versus strain plots for each deformation type are given in Figure~\ref{figure1}(h), at volume fraction $\phi=0.45$ and friction coefficient $\mu=1$.
For small strains $\dot{\gamma}t < 1$ we identify start-up regimes in which the viscosity increases with strain.
During this time, the number of direct particle-particle contacts increases with strain
and a flow-induced microstructure establishes~\cite{clark1980observation,Gadala-maria1980}.
As can be seen, we are able to achieve a
strain-independent region with a strain magnitude $\dot{\gamma}t = \mathcal{O}(1)$.
It is noted that the biaxial extension simulation is conducted using the output from the uniaxial extension as the initial condition. Thus the initial period presented for biaxial flow corresponds to a flow reversal rather than to a start-up from a quiescent state. Interestingly, a familiar characteristic surge in stresses~\cite{Gadala-maria1980} (hydrodynamic in origin) is observed at very small strains, indicative of placing closed particle contacts under tension as discussed by Refs~\cite{Ness2016b,lin2015hydrodynamic}.

In Figure~\ref{figure1}(i) we give the evolution of Trouton's ratio with strain, evaluated by rescaling the extensional viscosities by the shear viscosity at each strain increment. There are two interesting features to note. The most striking is that, for $\dot{\gamma} t>1$, the results are remarkably close to the Newtonian values. This suggests that the quasi-Newtonian character of overdamped suspensions extends beyond the linear scaling of shear stresses with shear rates. The result for planar extension matches that predicted by an independent simulation model~\cite{seto2017microstructure}. The second interesting feature is the surge in Trouton's ratio for each of the extensional flows, with a maximum at around $\dot{\gamma}t=0.5$. This indicates a faster microstructural evolution for extensional flows compared to shear flows. Such a finding is consistent with the form of the applied deformations, which see fluid elements move together/apart exponentially for extensional flows but linearly for shear flows.
The shaded regions in Figure~\ref{figure1}(i) represent the maximal and minimal values obtained over five independent simulation runs, indicating a very weak dependence on the initial configuration. Error bars are thus omitted from the following results and discussion.

\section{Evolution of the viscosity and Trouton's ratio with volume fraction}
\label{section_divergence}

\begin{figure*}
  \centering
  \includegraphics[trim = 0mm 60mm 0mm 0mm, clip,width=0.75\textwidth]{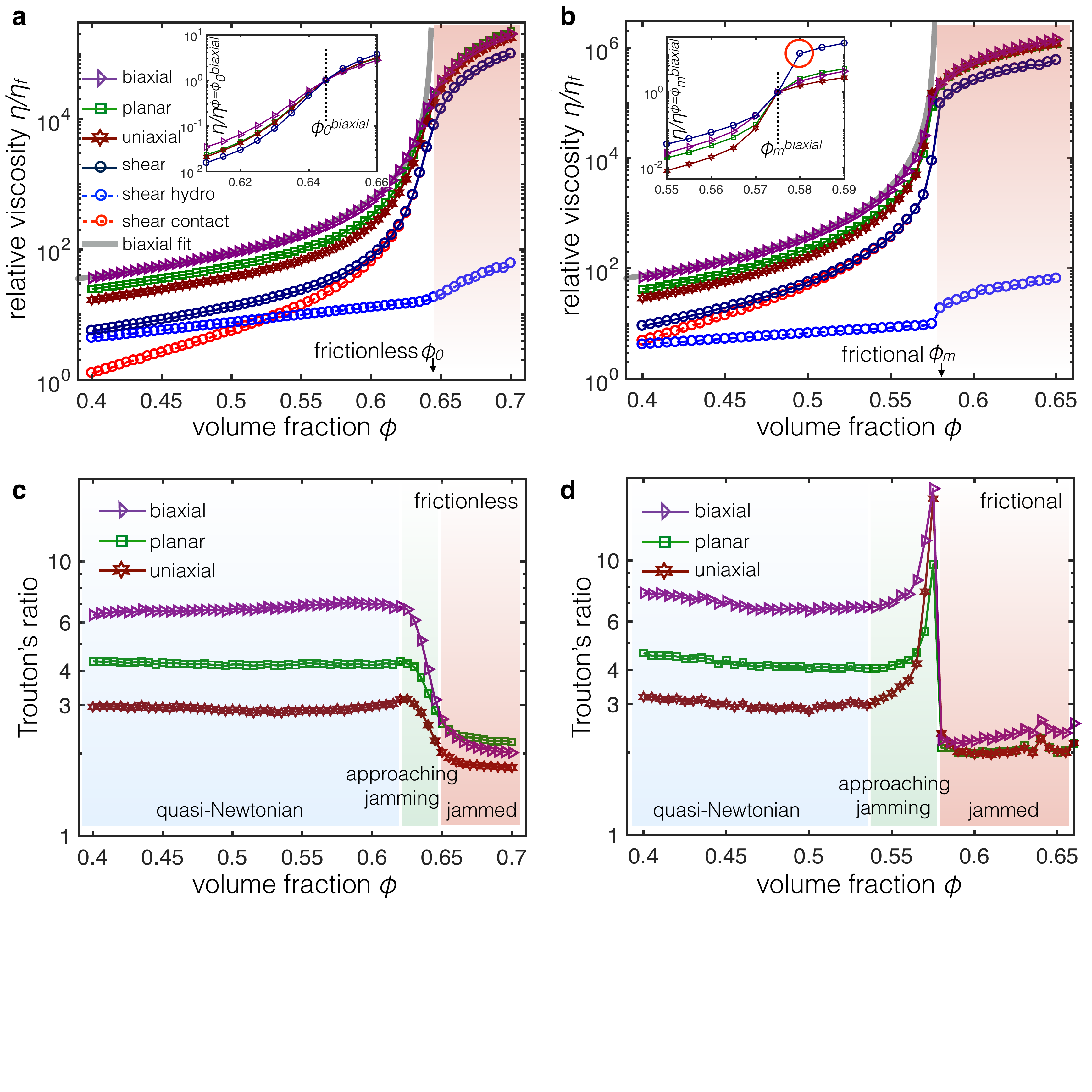}
     \caption{
Top: Divergence of the reduced viscosity $\eta/\eta_f$ as a function of volume fraction $\phi$ for shear, planar, uniaxial and biaxial flow with frictionless (a) and frictional (b) particles. Also shown are the relative hydrodynamic and contact contributions to the shear viscosity. Qualitatively equivalent results are obtained for the extensional flows (not shown).
Highlighted in the red shaded region in (a) and (b) are the `jammed' regions where the rheology is no longer expected to be viscous and thus a Newtonian Trouton's ratio is not expected.
Insets (a) and (b): Same data as (a) and (b) but focussing on the region near jamming where there is a discrepancy in jamming volume fraction. Plotted on the $y$-axes are the viscosities rescaled by their values at jamming (as measured under biaxial extension). Highlighted in Inset (b) in the red circle is the anomalous point for shear flow, which enters the jammed region at higher volume fractions than extensional flows.
Bottom: Evolution of Trouton's ratio with volume fraction $\phi$ for planar, uniaxial and biaxial flow with frictionless (c) and frictional (d) particles. Highlighted are regions where the ratio matches that of a Newtonian fluid (blue), where it deviates on the approach to jamming (green), and where it is fully jammed (red).
     }
 \label{figure2}
\end{figure*}

Presented in Figure~\ref{figure2} is the evolution of viscosity for frictionless (a) and frictional (b) particles
and the evolution of Trouton's ratio Tr for frictionless (c) and frictional (d) particles, with volume fraction $\phi$ for each deformation type.
In general, the viscosities for all flow types follow the form $\eta/\eta_f = \alpha(1 - \phi/\phi_c)^{-\beta}$,
with $\phi_c$ a generic `critical' volume fraction and $\beta$ a scaling parameter much discussed in the literature~\cite{krieger1959mechanism} and reported to be $\approx 2$ in shear flow experiments (see, for example,~\cite{Guy2015}).
Following conventional nomenclature~\cite{wyart2014discontinuous} we drop $\phi_c$ and label the frictionless and frictional jamming points $\phi_0$ and $\phi_m$, respectively, where $\phi_m$ depends on $\mu$.
We fitted such a form to our biaxial extension data and found $\alpha = 1.25$, $\beta=1.6$ and $\phi_0 = 0.644$ for frictionless and $\alpha = 1.1$, $\beta=2$ and $\phi_m = 0.575$ for frictional particles.
Also shown in (a) and (b) are the hydrodynamic and contact contributions to the shear viscosity which, when summed and complemented by the Stokes term $2\eta_f{\textbf E}^\infty$ lead to the total shear viscosity. We find that for frictional particles, contacts dominate even for $\phi<0.4$, while for frictionless particles contacts only become dominant at $\phi > 0.54$. Comparable behaviour of the contact and hydrodynamic viscosities is obtained for all of the flow types.

For $\phi>(\phi_0,\phi_m)$, the suspension enters a jammed state as indicated by the shaded red region in Figures~\ref{figure2}(a)-(d).
Here, flow is only possible through particle deformations and thus for strictly hard spheres, jamming represents flow arrest.
For the \emph{nearly}-hard spheres considered in this work,
we enter a high stress flowing regime in which an elasticity emerges corresponding to the stiffness of the particle-particle repulsion.
Such a region can only be observed experimentally when the particles are sufficiently soft, for example in emulsions~\cite{nordstrom2010}.
In any case, the flow in this region is not strictly viscous and thus is not expected to obey Newtonian Trouton's ratios.

The values of Tr presented in Figures~\ref{figure2}(c)-(d) demonstrate a remarkably broad range of volume fractions for which the flow appears to be approximately Newtonian,
persisting up to $\phi\approx0.62$ for frictionless and $\phi\approx0.54$ for frictional particles. Tr reaches between 7 and 8 under biaxial extension, but given the logarithmic scale over which the overall stresses are varying, we consider a discrepancy of $\approx25\%$ from the Newtonian value to be insubstantial.
Interestingly, our simulation result implies that there is very weak dependence of Trouton's ratio on particle-particle friction for $\phi<0.54$, despite the importance of the contact stress contribution and the dominant role of friction in setting the viscosity at these volume fractions.

An interesting disparity between frictionless and frictional particles emerges in the `approaching jamming' region,
for volume fractions $0.54<\phi<0.65$.
In the frictionless scenario, a narrow transition window of  $\Delta\phi\approx 0.02$ exists
in which Tr rapidly and monotonically switches from its low ($\phi<\phi_0$) plateau to a high ($\phi>\phi_0$) plateau.
The monotonicity suggests that each of the flowing states approach a common, deformation-type-independent, value of $\phi_0$.
The width of this `approaching jamming' region decreases with increasing particle stiffness
as the transition to jamming becomes sharper~(see Appendix).
By contrast, Tr for frictional particles begins to \emph{exceed} its Newtonian values around $\Delta\phi\approx 0.05$ below jamming $\phi_m$.
This suggests there is a window in which the
extensional viscosity of suspended frictional particles exceeds the shear viscosity by up to an order of magnitude.
In fact, we find (see Appendix) that this spike in the Trouton's ratio for frictional particles scales with the stiffness of the particles, strongly suggesting that at volume fractions in this region,
Tr actually represents a ratio between jammed and flowing states (rather than two stiffness-independent flowing states as is the case for frictionless particles, which show no such scaling) thus implying a discrepancy in $\phi_m$ for different flow types.
Returning to the viscosity divergence plotted in Figure~\ref{figure2}(b) Inset,
we verify that the surge in Tr corresponds to a mismatch in the frictional jamming volume fraction $\phi_m$ for different flow types, as highlighted by the red circle that indicates the entry to jamming for shear flow is shifted to the right with respect to extensional flows. The extensional viscosities tend to diverge at a common volume fraction that is approximately 0.005 below that for shear flow.
Based on this monotonicity and nonmonotonicity in Tr for frictionless and frictional particles respectively, we thus conclude that $\phi_m$ depends subtly upon the deformation type whereas $\phi_0$ does not. 
 (Though there appears to be a visual mismatch between $\phi_0$ values for different deformation types in Figure~\ref{figure2}a, the monotonicity of Tr (Figure~\ref{figure2}c) proves that the shift is only in the $y-$axis and not in $x$.)

It is also noted that there is weak $\phi$ dependence of Tr above $\phi_0$ ($\phi_m$) for frictionless (frictional) particles.
If we crudely take the rheology here to be quasistatic~\cite{chialvo2012bridging},
and thus dependent on the `shape' of the deformation tensor but not the relative magnitude of the deformation rate,
we can obtain a reasonable prediction of Tr above jamming.
Specifically, for planar, uniaxial and biaxial flows we obtain, respectively, Tr~$\sim4/2$, $\sim3/\sqrt{3}$ and $\sim6/(2\sqrt{3})$ above jamming,
regardless of particle-particle friction, corresponding to the representative viscosities rescaled by the magnitude of ${\textbf E}^\infty$ (see Table~\ref{table1}).

The deformation type dependence of $\phi_m$ suggests a clear route to intermittent jamming through changes in deformation type. For example, at a volume fraction of $\phi\approx 0.575$, a suspension of frictional particles is quasi-Newtonian under shear flow, but jammed under extensional flow. This poses a direct challenge to industrial processes that involve mixed flow, suggesting that a fluid element at fixed volume fraction might transiently jam and unjam dependent upon the instantaneous flow type to which it is subjected. Such an effect is not predicted for frictionless particles.

\section{Microstructural behaviour close to jamming}
\label{section_microstructure}

We observed a deformation-type-independent critical volume fraction $\phi_0$ for frictionless particles, but a deformation-type-dependent $\phi_m$ for frictional particles. This is consistent with earlier observations that frictional jamming, which occurs at $\phi_m$, shows protocol dependence and hysteresis. Flow arrest in frictional particles is thus often described as a fragile or shear-jamming transition that masks an underlying true jamming transition which occurs at $\phi_0$ (with $\phi_0>\phi_m$)~\cite{bi2011jamming}.

When frictional forces are large, percolating chains of stable but fragile particle-particle contacts
can permit jamming with considerable anisotropy at volume fractions below $\phi_0$
(see, for example, Refs~~\cite{otsuki2011critical,grob2014jamming,bi2011jamming}).
In such systems, experiments show hysteretic effects whereby the material initially jams at some low packing fraction (similar to our $\phi_m$ here)
but upon further perturbations it consolidates and approaches $\phi_0$~\cite{bandi2013fragility}.
No such hysteresis is observed at frictionless jamming~\cite{ciamarra2011jamming},
which thus occurs when the material reaches an isotropic (or, at least, more isotropic, see~\citet{baity2017emergent}) packed state.
Our suspension of frictionless particles might thus reach jamming at an isotropic configuration that is not protocol (\emph{i.e.} deformation type) dependent,
whereas the frictional particles reach jamming when their dynamically evolving force chains are able to percolate the system and permit an anisotropic jammed state, which is necessarily protocol dependent.

To test whether this description is suitable for explaining our observed divergence of Tr for frictional
(but not frictionless) particles, we consider the microstructural anisotropy at the critical volume fraction.
To do this, we consult a familiar form of fabric tensor defined as $A_{ij} = \langle n_{i}n_{j} \rangle - \frac{1}{3}\delta_{ij}$~\cite{sun2011constitutive},
where $n_i$ is a particle-particle unit vector and angular brackets represent
an average over all particles that are in mechanical contact (defined when $|{\bm r}|<(a_1 + a_2)$).
For a large, isotropic sample, one obtains $A_{ij}\to0$.
We use scalar representations of the fabric, $A$,
corresponding to the viscosity definitions given in Table~\ref{table1},
for shear $A\coloneqq A_{xy}$, planar  $A\coloneqq A_{yy}-A_{xx}$, uniaxial $A\coloneqq A_{zz}-A_{xx}$ and biaxial  $A\coloneqq A_{xx}-A_{zz}$ deformations.

\begin{figure*}
  \centering
  \includegraphics[trim = 0mm 0mm 40mm 0mm, clip,width=1\textwidth]{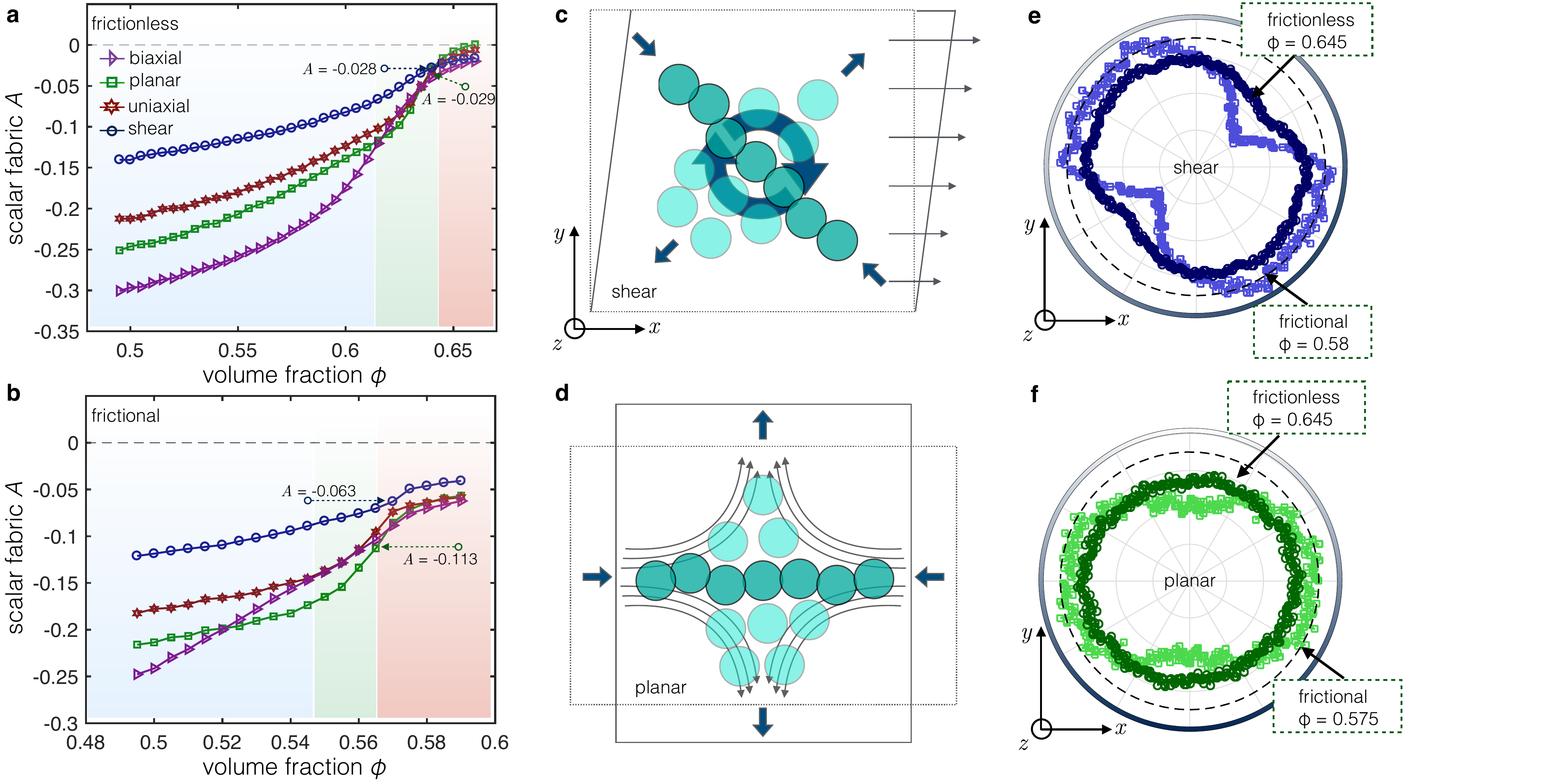}
     \caption{
     Behaviour of the microstructural fabric close to the critical volume fraction for each deformation type explored in this work. Shown are the scalar fabric $A$ for
     (a) frictionless particles and
     (b) frictional particles.
     Dashed arrows show exemplary values of $A$ at jamming.
     In (c) and (d) we draw schematic illustrations of microstructural fabric for shear flow and planar extension, respectively. Dark shading corresponds to load-bearing contacts while light shading corresponds to `spectator' particles~\cite{Cates1998}. Straight blue arrows indicate compressive axes; curved blue arrows indicate rotational component; grey arrows indicate streamlines corresponding to affine flow.
     Shown in (e) and (f) are radial distributions of particle-particle contacts (defined when $|{\bm r}|<(a_1 + a_2)$) at the critical volume fraction, projected onto $xy$ for shear and planar flows. The shape of the distribution reflects the values of $A$ (dashed black lines show the result when our algorithm is run using $5\times10^5$ random points distributed uniformly over a spherical surface). Suspensions of frictionless particles jam with $A$ less than half the frictional value, and thus the distribution is more circular.
     }
 \label{figure3}
\end{figure*}

Fabric data are presented for frictionless and frictional particles in Figures~\ref{figure3}(a) and (b) respectively.
In all cases negative $A$ corresponds to a preferential orientation of contacts along the compressive flow axis,
which we illustrate schematically in Figures~\ref{figure3}(c) and (d) for shear and planar extension deformations, respectively.
In Figures~\ref{figure3}(e) and (f) we plot the distribution of the vector $n_i$ projected onto the $xy$-plane for shear and planar flow.
There is always an alignment of contacts along the compressive axes,
regardless of $\phi$, friction and deformation type.
We find that $A\to0$ as $\phi$ increases, indicating that the microstructure generally becomes more isotropic as jamming is approached.
Crucially, it is observed that there is a strong disparity in the values of $A$ measured at the jamming point when comparing frictionless and frictional particles.
Frictionless particles jam when $A$ is closer to 0 (indicative examples are $A=-0.028$ for shear and $A=-0.029$ for planar deformations at $\phi_0$, indicated in Figure~\ref{figure3}a) indicating that flow-arrest is achieved with a more isotropic microstructure than in frictional flows,
which have $A=-0.063$ and $A=-0.113$ respectively at their respective $\phi_m$ (Figure~\ref{figure3}b).
This finding is also apparent in the radial distributions shown in Figures~\ref{figure3}(e)-(f).
These show a more anisotropic distribution of contact forces at the jamming volume fraction for frictional compared to frictionless particles,
with a surplus of particle contacts along the NW-SE axis under shear flow and the E-W axis under planar flow.
In contrast, the profiles for frictionless particles are,
while not perfectly circular, rather more uniform.
Moreover, there is little deformation type dependence in the value of $A$ at $\phi_0$ for frictionless particles, suggesting that, although the definitions of $A$ vary with each case, jamming occurs with a similarly isotropic structure in each case.
By contrast, there is quite some variation in $A$ at $\phi_m$ for frictional particles, again emphasising the dependence upon deformation type.

Frictionless particles only jam when their arrangement is nearly isotropic, so it doesn't matter what type of deformation we apply; frictional particles can jam in an anisotropic state, so it matters how we deform them up to this point.
 We thus conclude that frictionless particles have constant Tr all the way to $\phi_0$ because different deformation types share this critical volume fraction; frictional particles have a deformation type dependent $\phi_m$ which is highest for shear flows, meaning Tr diverges between e.g. $\phi_m^\text{uniaxial}$ and $\phi_m^\text{shear}$. 
 
\section{Mapping the extensional deformations onto viscous number rheology}
\label{section_viscous}

\begin{figure*}
  \centering
  \includegraphics[trim = 0mm 220mm 0mm 0mm, clip,width=0.75\textwidth]{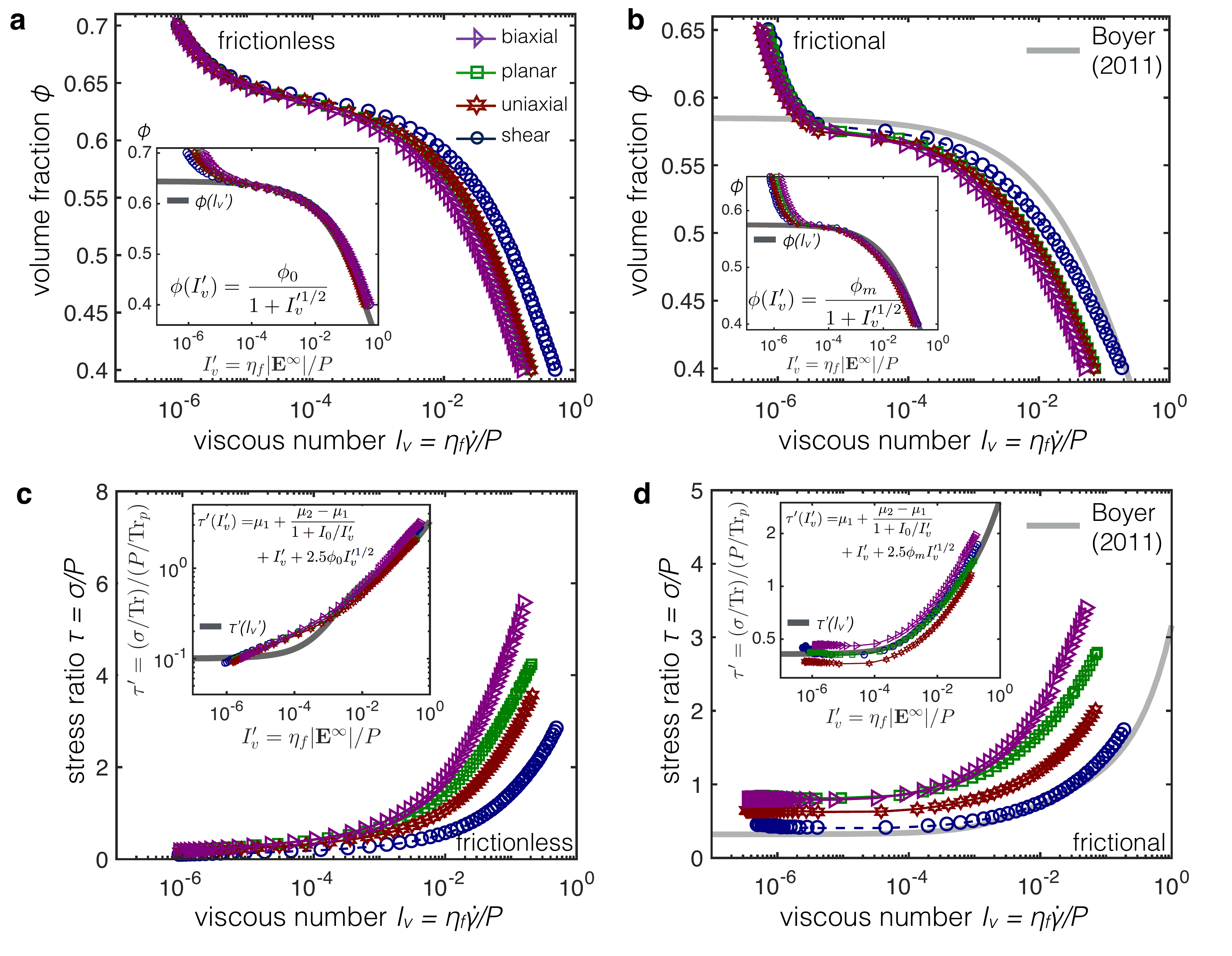}
     \caption{Viscous number rheology for shear and extensional flows.
     Shown are the volume fraction as a function of viscous number for frictionless (a) and frictional (b) particles,
     and the stress ratio as a function of viscous number for frictionless (c) and frictional (d) particles.
     We redefined the viscous number replacing $\dot{\gamma}$ with $|{\textbf E}^\infty|$
     and redefined the stress ratio by rescaling $\sigma$ with a Newtonian Trouton's ratio for each flow type.
     We thus arrive at the collapsed plots of volume fraction as a function of viscous number for frictionless ((a), Inset) and frictional particles ((b), Inset)
     and stress ratio as a function of viscous number for frictionless ((c), Inset) and frictional particles ((d), Inset). We provide fits to the Inset data according to the expressions given therein, with the parameters given in the main text.
     Also shown in (b) and (d) are the predictions given by~\citet{boyer2011unifying} based on shear flow experiments.
     }
 \label{figure4}
\end{figure*}

We finally verify that the numerical model described herein predicts flow behaviour under shear and extension
that qualitatively follows the viscous number rheology framework proposed by~\citet{boyer2011unifying} very well.
The viscous number is defined as $I_v = \eta_f\dot{\gamma}/P$ (for suspending fluid viscosity $\eta_f$, deformation rate $\dot{\gamma}$ and pressure $P$) and works as an analogue of the inertial number used in dry granular material modelling~\cite{jop2006constitutive}.
In the athermal, non-inertial limit, the rheological state of a suspension can be uniquely defined using two functions that relate the volume fraction $\phi$ and the stress ratio $\tau = \sigma/P$ to the viscous number $I_v$.
The stress ratio, which is ordinarily taken as the ratio between the shear stress and mean normal stress (i.e. the pressure $P$),
is defined in this work according to the viscosity definitions given in Table~\ref{table1}.
Specifically, we replace the shear stress with a generic stress $\sigma$ given by $\sigma\coloneqq\sigma_{xy}$ for shear,
$\sigma \coloneqq \sigma_{yy}-\sigma_{xx}$ for planar, $\sigma \coloneqq \sigma_{zz}-\sigma_{xx}$ for uniaxial and $\sigma \coloneqq \sigma_{xx}-\sigma_{zz}$ for biaxial deformations.
For each flow type, the pressure is taken simply as $P = -\frac{1}{3}\sum_{i=x,y,z} \sigma_{ii}$.
The functions $\phi(I_v)$ and $\tau(I_v)$ are presented in Figure~\ref{figure4}.
Crucially, qualitatively consistent behaviour is observed for both shear and extensional flows and for both frictionless and frictional particles.
Comparing frictionless and frictional cases quantitatively, we find discrepancies in the critical $\phi$, as discussed above,
as well as discrepancies in the limiting $\tau$ at low $I_v$, which has been discussed earlier by~\citet{da2005rheophysics}.
We also show in Figure~\ref{figure4}(b) and (d) the predictions based on the model proposed by~\citet{boyer2011unifying}, for which they give parameters appropriate for frictional particles (we use their parameters here).
Sources of discrepancy between the present result and the model prediction are variations in polydispersity
(which alter the numerical value of the critical volume fraction $\phi$ measured when $I_v\to 0$),
variations in particle-particle friction coefficient (which alter the numerical value of the limiting stress ratio $\sigma/P$ measured as $I_v\to0$)
and variations in particle hardness (which alter the critical viscous number at which volume fractions may exceed the critical volume fraction).

Considering $\phi(I_v)$ in Figures~\ref{figure4}(a) and (b), we find some discrepancy in the quantitative results for different flow types below the critical volume fraction.
Interestingly, we find that a convincing collapse of the data in this region is obtained if we redefine $I_v$ based on the magnitude of the deformation rate tensor,
that is replacing $\dot{\gamma}$ with $|{\mathbf E}^\infty|$ to give $I_v' = \eta_f|{\textbf E}^\infty|/P$, Figure~\ref{figure4}(a) and (b) [Inset].
This result implies that an alternative Trouton's ratio ($\text{Tr}_p$) may be defined for the mean normal stresses,
taking values that correspond approximately to $|{\textbf E}^\infty|$ for each flow type.
Comparing $\tau(I_v)$ in Figures~\ref{figure4}(c) and (d),
we similarly find a qualitative match for all flow types, but a quantitative discrepancy.
Since we have demonstrated satisfactory correspondence to Newtonian Tr for a broad range of $\phi$,
as well as a convincing collapse of $\phi(I_v)$ with $|{\mathbf E}^\infty|$ that implies an equivalent $\text{Tr}_p$,
we crudely define a rescaled stress ratio as $\tau' = (\sigma/\mathrm{Tr})/(P/\mathrm{Tr}_p)$.
Using $\tau'$ and $I_v'$ as defined above, we again are able to collapse the data, Figure~\ref{figure4}(c) and (d) [Inset].
The collapse is particularly convincing for frictionless particles and still rather good for frictional particles.

This result demonstrates that, provided the stresses are rescaled appropriately by their Trouton's ratios
(which we have shown can be considered as Newtonian for a broad range of $\phi$),
the viscous number rheology framework proposed by~\citet{boyer2011unifying}
can predict the rheology for all of the deformations considered in this work with a single set of parameters.
We provide examples of such a fitting for the frictionless case (Figures~\ref{figure4}(a) and (c) (Inset)), using
$
\phi(I_v') = \phi_0/(1 + I_v'^{1/2})
$
and
$
\tau'(I_v') = \mu_1 + (\mu_2-\mu_1)/(1 + I_0/I_v') + I_v' + 2.5\phi_0I_v'^{1/2}
$
with $\phi_0=0.644$, $\mu_1=0.1$, and the parameters $\mu_2=0.7$ and $I_0=0.005$ following~\citet{boyer2011unifying}.
Similarly for the frictional case (Figures~\ref{figure4}(b) and (d) (Inset)), with
$
\phi(I_v') = \phi_m/(1 + I_v'^{1/2})
$
and
$
\tau'(I_v') = \mu_1 + (\mu_2-\mu_1)/(1 + I_0/I_v') + I_v' + 2.5\phi_mI_v'^{1/2}
$
with $\phi_m=0.575$ and the parameters $\mu_1=0.32$, $\mu_2=0.7$ and $I_0=0.005$ following~\citet{boyer2011unifying}.
Since the necessary stress rescalings derive directly from the relationships between the rate of strain tensors defined above,
and according to Newtonian rheology (at least for volume fractions up to slightly below jamming),
we can characterise the rheology of athermal, noninertial particle suspensions in any of the studied flows based on the rate-independent formulation~\cite{boyer2011unifying}. Interestingly, the log-linear axes in Figures~\ref{figure4}(c) and (d) Inset reveal a potential mismatch in the functional form of $\tau'$ for frictionless and frictional particles on the approach to jamming. We expect that this does not derive from the effects of polydispersity or particle hardness mentioned above, but rather represents a qualitative difference in the nature of the stresses at flow arrest when contacts are sliding or rolling. The asymptotic behaviour of $\tau'(I_v')$ for frictionless particles has been previously demonstrated in the absence of hydrodynamic interactions~\cite{lerner2012unified}, and further analyses based on the current model are deferred to future work.

\section{Conclusion}

We have thus shown that for a broad range of volume fractions the underlying extensional rheology of dense suspensions can be described simply by a Newtonian Trouton's ratio.
This leads to a good agreement with viscous number rheology, provided the stresses are rescaled appropriately by the Trouton's ratio, which is available \emph{a priori} from the known deformation tensor.
For suspensions of frictionless particles, our model predicts no flow-type dependence on the critical volume fraction for jamming $\phi_0$ and consequently the Trouton's ratios are fixed up to $\phi\approx0.63$.
This result is relevant for athermal suspensions with normal repulsive interactions between particles, for example emulsions and silica suspensions below shear thickening.
By contrast, a disparity in jamming volume fractions $\phi_m$ for different deformations emerges for frictional particles,
suggesting that mixed flows with shear and extensional components might jam and unjam at fixed volume fraction \emph{and} stress, simply due to changes in the deformation.
This is relevant for suspensions of large granular particles of the type described under the framework of~\citet{boyer2011unifying}, and also for silica suspensions above the onset of shear thickening.

It would be interesting to determine whether, in practice, chaotic flow or even oscillating flows of the type described by~\citet{pine_chaos_2005} that can eliminate particle-particle contacts might serve to inhibit the role of load-bearing force chains and thus extend the range of volume fractions that exhibit Newtonian Trouton's ratios even for frictional particles.
Achieving a general description of extensional rheology is relevant to numerous applications that involve mixed flows of dense suspensions, notably in footstuffs~\cite{huang1993measurement}, ceramic paste extrusion~\cite{benbow1993paste,ness2017linking} and calcium phosphate injections for bone replacement treatments~\cite{zhang2014calcium}. In addition, dense suspensions are emerging as a useful material for energy dissipation during impacts, for which both biaxial~\cite{gurgen2017stab} and uniaxial~\cite{ballantynerate} configurations are relevant.

\section{Acknowledgement}

CN acknowledges the Maudslay-Butler Research Fellowship at Pembroke College, Cambridge for financial support.
We thank Jin Sun and Zeynep Karatza for useful discussions and Ranga Radhakrishnan for sharing his derivation of the lubrication forces~\cite{Ranga2017}.
A video showing a typical deformation and the scripts needed to reproduce it are given at \url{https://doi.org/10.17863/CAM.13415}.

\appendix{
\setcounter{figure}{0} \renewcommand{\thefigure}{A\arabic{figure}}
\section{Scalar resistances for hydrodynamic lubrication forces}
\label{app:app1}

The scalar resistances used in the hydrodynamic force model described in Section \ref{section_model}  follow those presented by Kim and Karrila~\cite{kim1991microhydrodynamics} and are given (for $\beta = a_2/a_1$) by
\begin{subequations}
\begin{equation}
X^A_{11} = 6\pi a_1\left(\frac{2\beta^2}{(1+\beta)^3}\frac{1}{\xi} + \frac{\beta(1+7\beta+\beta^2)}{(5(1+\beta)^3)}\ln\left(\frac{1}{\xi}\right) \right) \text{,}
\end{equation}
\begin{equation}
Y^A_{11} = 6\pi a_1\left(\frac{4\beta(2+\beta+2\beta^2)}{15(1+\beta)^3}\ln\left(\frac{1}{\xi}\right) \right) \text{,}
\end{equation}
\begin{equation}
Y^B_{11} = -4\pi a_1^2\left(\frac{\beta(4+\beta)}{5(1+\beta)^2}\ln\left(\frac{1}{\xi}\right)\right) \text{,}
\end{equation}
\begin{equation}
Y^B_{21} = -4\pi a_2^2\left(\frac{\beta^{-1}(4+\beta^{-1})}{5(1+\beta^{-1})^2}\ln\left(\frac{1}{\xi}\right)\right) \text{,}
\end{equation}
\begin{equation}
Y^C_{11} = 8\pi a_1^3\left(\frac{2\beta}{5(1+\beta)}\ln\left(\frac{1}{\xi}\right)\right) \text{,}
\end{equation}
\begin{equation}
Y^C_{12} = 8\pi a_1^3\left(\frac{\beta}{10(1+\beta)}\ln\left(\frac{1}{\xi}\right)\right) \text{.}
\end{equation}
\end{subequations}

\section{Check that the simulation result isn't affected by finite-size effects}

Using simple shear as a test case, we simulate various periodic box sizes (i.e. particle numbers) to check that there are no finite size effects.
For simulations with $N>3000$, we find rather convincing system size independence. Thus we conclude that the results presented in this work, which all have N$>10000$, are not influenced by system size.

\begin{figure}[b]
  \centering
  \includegraphics[width=0.35\textwidth]{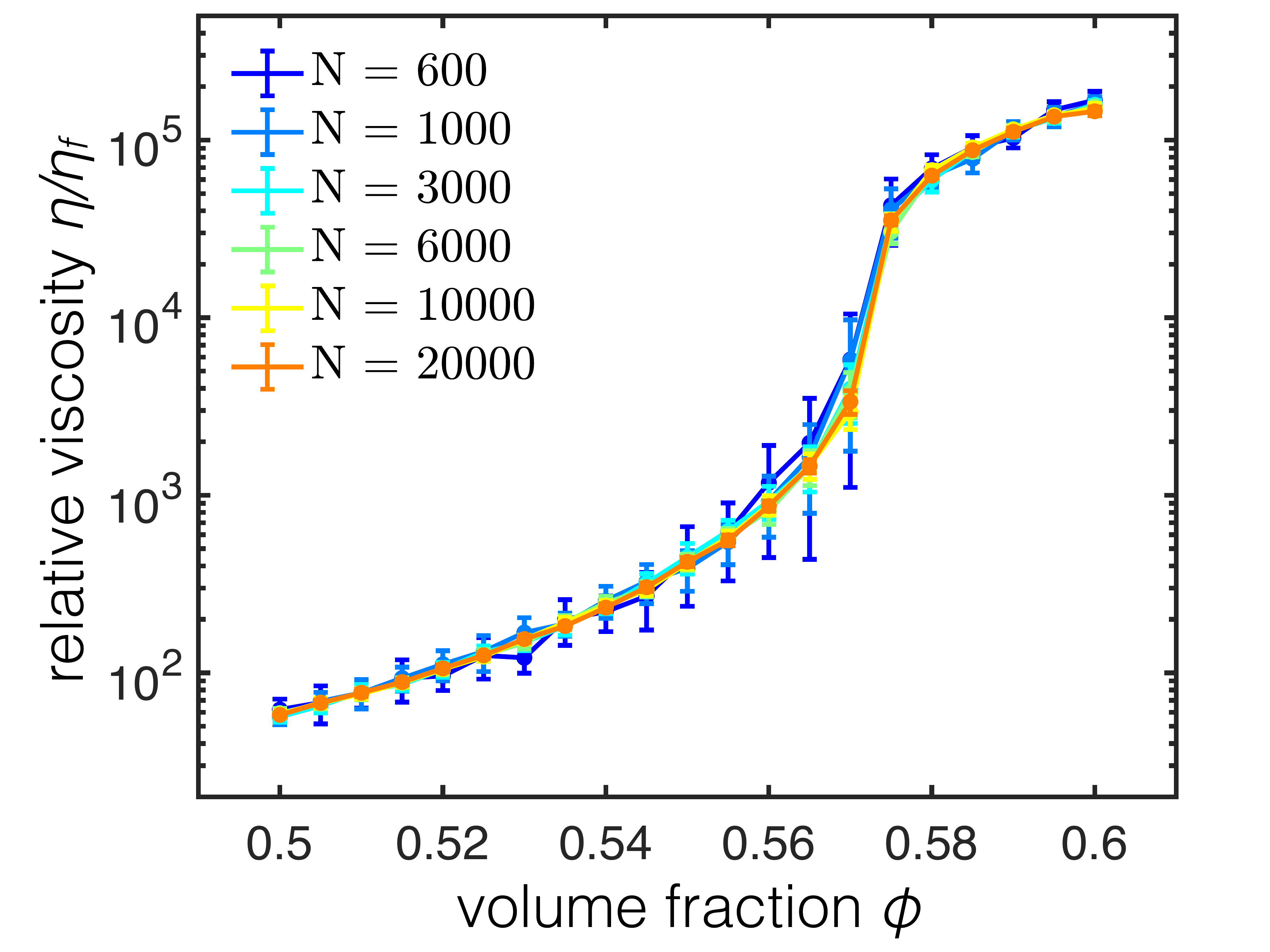}
     \caption{
Divergence of simple shear viscosity for frictional particles ($\mu = 1$) with volume fraction for several different system sizes, measured in terms of total particle number N.
     }
 \label{figurea2}
\end{figure}

\section{Demonstration of simulation breaking down for large uniaxial strains}

As discussed in the main text, our extensional flow simulations do not allow arbitrarily large deformations, but rather are limited by the shrinking length of the compressive axes. We tested the maximal strain that can be reasonably achieved under uniaxial extension by deforming the box until the measured viscosity shows unphysical behaviour, Figure~\ref{figurea2}. For both system sizes considered, we are able to obtain a strain independent viscosity in the strain window $1\to4.5$. We thus constrain the averaging window for all extensional flow simulations considered in this work to that range of strains.

\begin{figure}
  \centering
  \includegraphics[width=0.4\textwidth]{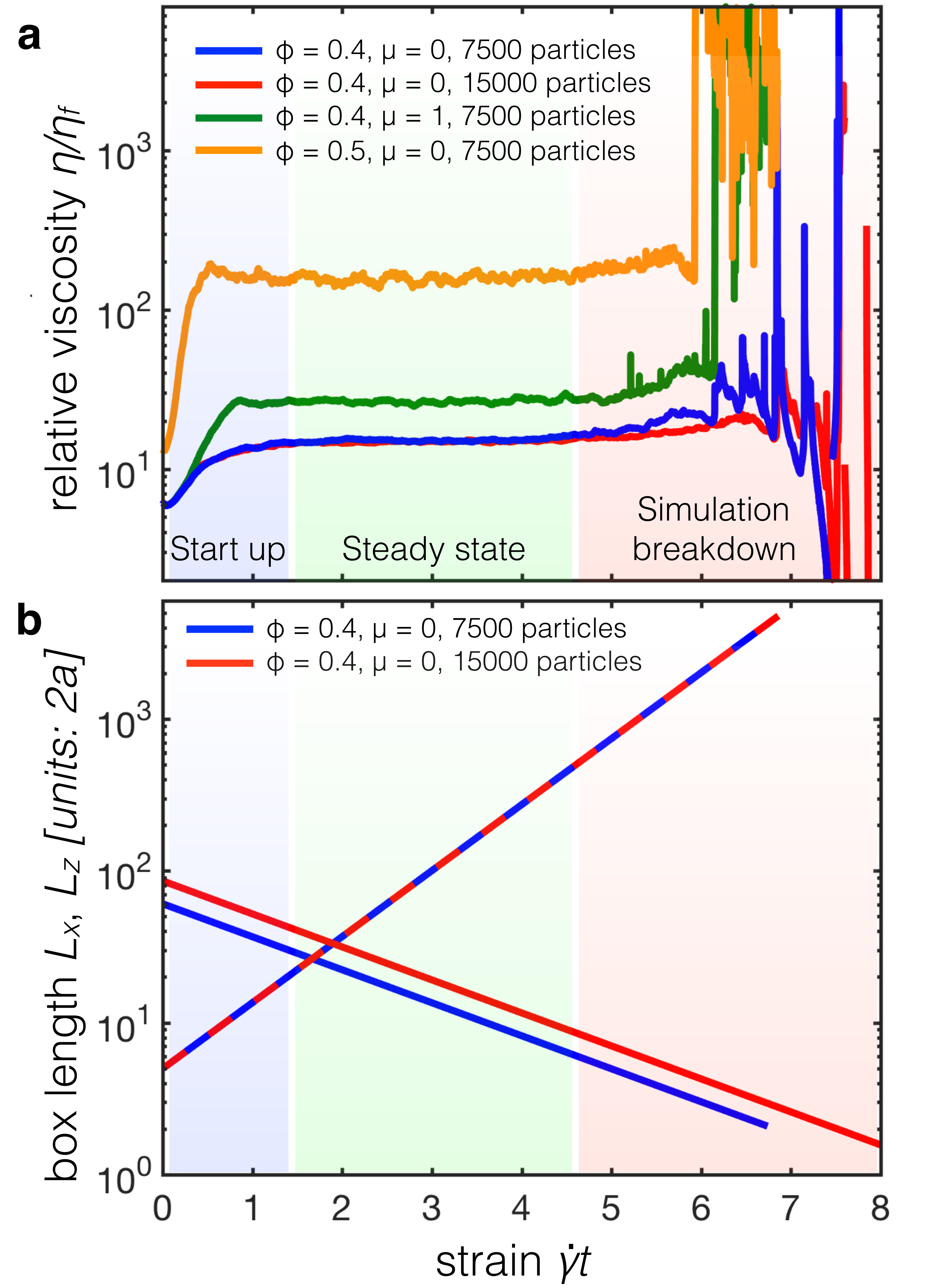}
     \caption{
(a) Viscosity versus strain plot for uniaxial deformation with a small (N=7500) and large (N=15000) simulation box.
(b) Variation of box dimensions $L_x$ and $L_z$ with strain.
     }
 \label{figurea2}
\end{figure}

\section{Role of particle stiffness}
To confirm that the spike in Trouton's ratio observed for frictional particles does indeed represent a ratio between a flowing and a jammed state, we repeated the simulations using particles with increased stiffness. Since the stresses in the flowing states are roughly independent of particle stiffness (since we are already near the hard particle limit) while the jammed state stresses scale with $k_n/a$~\cite{chialvo2012bridging}, we find that the magnitude of the spike in Trouton's ratio for frictional particles at $\phi=0.575$ scales with the particle stiffness,~Figure~\ref{figurea3}.
\begin{figure}
  \centering
  \includegraphics[trim = 0mm 0mm 35mm 0mm, clip,width=0.4\textwidth]{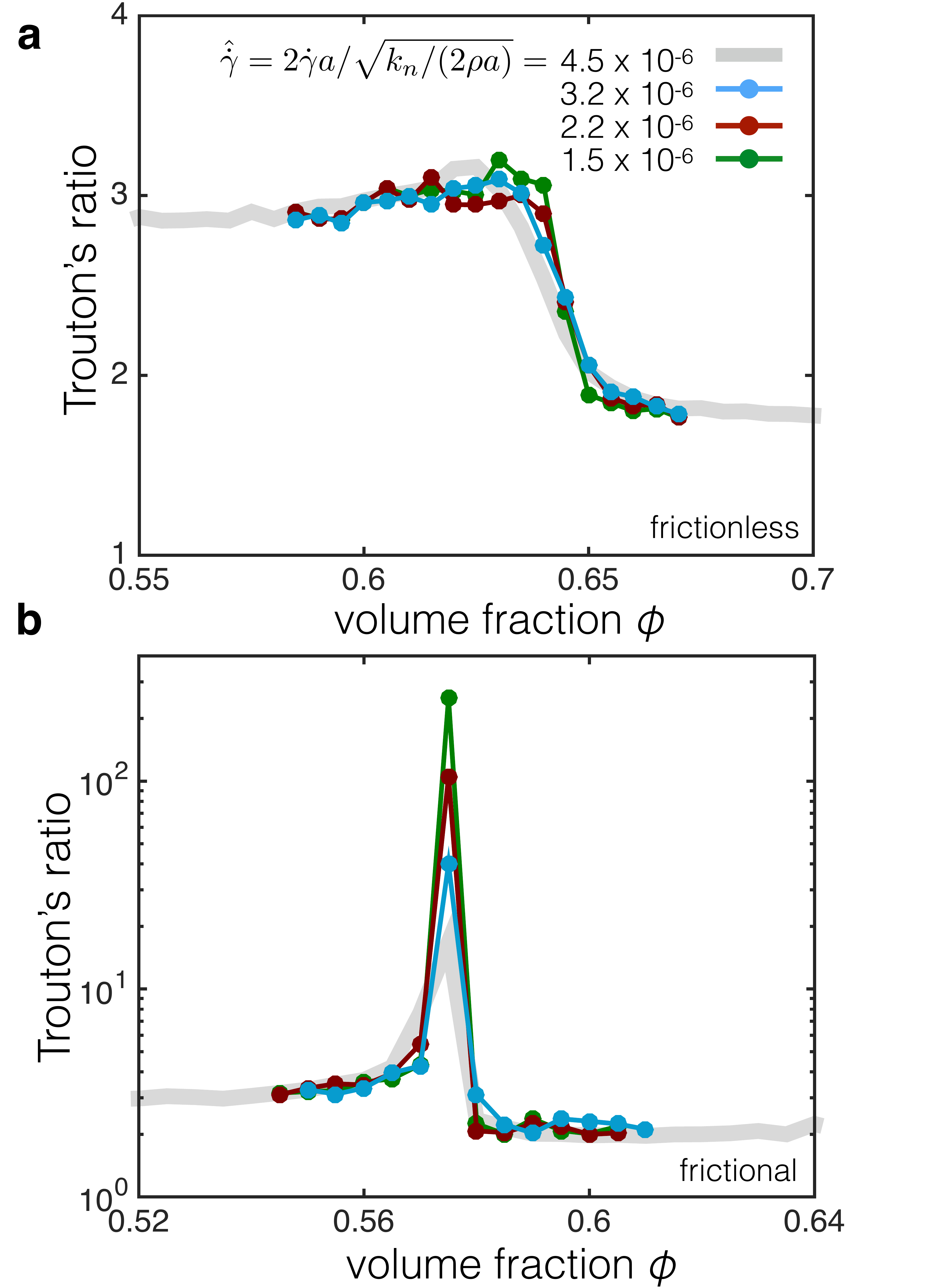}
     \caption{
(a) Trouton's ratio versus volume fraction for uniaxial deformation and fricitonless particles with increasing particle stiffness $k_n$, $k_t$.
(b) Trouton's ratio versus volume fraction for uniaxial deformation and fricitonal particles with increasing particle stiffness $k_n$, $k_t$.
     }
 \label{figurea3}
\end{figure}

}

\bibliography{library.bib}

\end{document}